\input{aipcheck}

\documentclass[
    ,final            
  ]
  {aipproc}

\layoutstyle{6x9}

\def\aap{\emph{A\&A}\ }

\def\apj{\emph{ApJ}\ }
\def\apjs{\emph{ApJS}\ }

\def\mnras{\emph{MNRAS}\ }
\def\nphysa{\emph{Nucl. Phys. A}\ }
\def\nat{\emph{Nature}\ }

\def\lsim{\mathrel{\rlap{\lower 4pt \hbox{\hskip 1pt $\sim$}}\raise 1pt
\hbox {$<$}}} 
\def\gsim{\mathrel{\rlap{\lower 4pt \hbox{\hskip 1pt $\sim$}}\raise 1pt
\hbox {$>$}}}
\newcommand{\etal}{et~al.}

\newcommand{\eg}{e.g., }
\newcommand{\ie}{i.e., }
\newcommand{\Msun}{M_{\odot}}

\newcommand{\ergs}{erg~s$^{-1}$}

\newcommand{\Nifs}{$^{56}$Ni}

\newcommand{\Ed}{\dot{E}_{\rm dep}}

\newcommand{\Edep}{\dot{E}_{\rm dep,51}}
\newcommand{\Et}{{E}_{\rm dep}}

\newcommand{\thj}{\theta_{\rm jet}}
\newcommand{\fth}{f_{\rm th}}
\newcommand{\Gj}{\Gamma_{\rm jet}}

\newcommand{\Min}{M_0}
\newcommand{\Rin}{R_0}
\newcommand{\Mfe}{M{\rm (Fe)}}

\newcommand{\Mjet}{M_{\rm jet}}
\newcommand{\Mbh}{M_{\rm rem}}
\begin{document}

\title{Supernova Nucleosynthesis and Extremely Metal-Poor Stars}


\author{Nozomu~Tominaga}{
  address={Department of Astronomy, University of Tokyo, Bunkyo, Tokyo
113-0033, Japan}
}

\author{Hideyuki~Umeda}{
  address={Department of Astronomy, University of Tokyo, Bunkyo, Tokyo
113-0033, Japan}
}

\author{Keiichi~Maeda}{
  address={Institute for the Physics and Mathematics of the Universe,
  University of Tokyo, Kashiwa, Chiba, 277-8582, Japan}
}

\author{Nobuyuki~Iwamoto}{
  address={Nuclear Data Center, Nuclear Science and Engineering
Directorate, Japan Atomic Energy Agency, Tokai, Ibaraki 319-1195, Japan}
}

\author{Ken'ichi~Nomoto}{
  address={Department of Astronomy, University of Tokyo, Bunkyo, Tokyo
113-0033, Japan},
  altaddress={Institute for the Physics and Mathematics of the Universe,
  University of Tokyo, Kashiwa, Chiba, 277-8582, Japan}
}

\begin{abstract}
 We investigate hydrodynamical and nucleosynthetic properties of the
 jet-induced explosion of a population III $40\Msun$ star and compare
 the abundance patterns of the yields with those of the metal-poor
 stars. We conclude that (1) the ejection of Fe-peak products and the fallback of unprocessed
 materials can account for the abundance patterns of the extremely
 metal-poor (EMP) stars and that (2) the jet-induced explosion with different energy
 deposition rates can explain the diversity of the abundance patterns
 of the metal-poor stars. Furthermore, the abundance
 distribution after the explosion and the angular dependence of the
 yield are shown for the models with high and low energy deposition rates
 $\Ed=120\times10^{51}$\ergs\ and $1.5\times10^{51}$\ergs. We also
 find that the peculiar abundance pattern of a
 Si-deficient metal-poor star HE~1424--0241 can be reproduced by the
 angle-delimited yield for $\theta=30^\circ-35^\circ$ of the model with
 $\Ed=120\times10^{51}$ \ergs. 
\end{abstract}

\maketitle


\section{INTRODUCTION}
\label{sec:introduction}

In the early universe, the enrichment due to a single supernova (SN) dominates the
preexisting metal contents (\eg \cite{aud95}).
A shock wave compresses the SN ejecta consisting of metals, 
\eg C, O, Mg, Si, and Fe, and the circumstellar materials
consisting of H and He. The abundance pattern of the enriched gas 
reflects nucleosynthesis in the SN. The compression will initiate a 
star formation, called a SN-induced star
formation (\eg \cite{cio88}); the next-generation stars will be
formed from the enriched gases. Thus the abundances of the
next-generation stars show a trace of nucleosynthesis in the population (Pop) III
SN. The low-mass ($\sim1\Msun$) stars among them have long lifetimes,
survive until present days, and might be observed as metal-poor
stars. Therefore, the metal-poor stars can make a constraint on
the nucleosynthesis yield of the Pop III SN.

The abundance patterns of extremely metal-poor (EMP) stars with [Fe/H]
$<-3$\footnote[1]{Here [A/B] 
$=\log_{10}(N_{\rm A}/N_{\rm B})-\log_{10}(N_{\rm A}/N_{\rm B})_\odot$, 
where the subscript $\odot$ refers to the solar value and $N_{\rm A}$
and $N_{\rm B}$ are the abundances of elements A and B, respectively.} 
suggest that aspherical SN explosions took place in the early universe. The
C-enhanced type of the EMP stars has been well explained by a faint
SN \cite{ume05,iwa05,nom06,tom07b}, except for their large Co/Fe and
Zn/Fe ratios (\eg \cite{dep02,bes05}). The enhancement of Co and Zn in low
metallicity stars requires explosive nucleosynthesis under high
entropy. In a {\sl spherical} model, a high entropy explosion corresponds to a
high energy explosion that inevitably synthesizes a large amount of
\Nifs(Fe) and leads a bright SN. Thus, it was suggested that some faint
SNe were associated with a narrow jet within which a high entropy region
is confined \cite{ume05}. 

We present hydrodynamical and nucleosynthetic models of
the jet-induced explosions of a Pop III 40 $\Msun$ star and show 
that the jet-induced SN explosions are responsible for the
formation of the EMP stars. Further, we investigate the angular
dependence of the yield of the jet-induced SNe.

\section{MODELS}
\label{sec:model}

We investigate a jet-induced SN explosion of a Pop III $40\Msun$ star
\citep{ume05,tom07b} by means of a two-dimensional relativistic Eulerian
hydrodynamic and nucleosynthesis calculation with the gravity
\cite{tom07a,tom07c,tom07d}. The nucleosynthesis calculation is
performed as a post-processing with the reaction
network including 280 isotopes up to $^{79}$Br (\cite{hix96,hix99}, see Table~1 in \cite{ume05}). 
The thermodynamic histories are traced by maker particles representing 
Lagrangian mass elements (\eg \cite{hac90,mae03b}). 

The explosion mechanism of gamma-ray burst (GRB)-associated SNe is still under debate (\eg a neutrino
annihilation, \cite{woo93}; and a magneto-rotation,
\cite{bro00}). Thus, we do not consider how the jet is
launched, but we deal the jet parametrically with the following five
parameters \cite{tom07d}: energy deposition rate ($\Ed$),
total deposited energy ($\Et$), initial half angle of the jets ($\thj$),
initial Lorentz factor ($\Gj$), and the ratio of thermal to total
deposited energies ($\fth$).

Jets are injected at a radius $\Rin \sim 900$ km, corresponding to an
enclosed mass of $\Min \sim 1.4 \Msun$, and the jet propagation is
followed until a homologously expanding structure is reached ($v\propto r$). 
Then, the ejected mass elements are identified from whether their radial
velocities exceed the escape velocities at their position. 

We investigate the dependence of nucleosynthesis outcome on $\Ed$ for a
range of $\Edep\equiv\Ed/10^{51}{\rm ergs\,s^{-1}}=0.3-1500$ and in
particular show the abundance patterns of the yields of two models; (A)
a model with $\Edep=120$, (B) a model with $\Edep=1.5$. Here, we fix the
other parameters as $\Et=1.5\times10^{52}$ergs, $\thj=15^\circ$, 
$\Gj=100$, and $\fth=10^{-3}$ in all models. 
The mass of jets is $\Mjet\sim8\times10^{-5}\Msun$. The model
parameters, the ejected Fe mass [$\Mfe$], and the central remnant mass
($\Mbh$) are summarized in Table~\ref{tab:2Drelmodel}.

\begin{table}
\begin{tabular}{c|ccccccc|cc}
\hline
   \tablehead{1}{c}{}{Name}
 & \tablehead{1}{c}{}{$\Min$}
 & \tablehead{1}{c}{}{$\Rin$}
 & \tablehead{1}{c}{}{$\Ed$}
 & \tablehead{1}{c}{}{$\Et$}
 & \tablehead{1}{c}{}{$\thj$}
 & \tablehead{1}{c}{}{$\Gj$}
 & \tablehead{1}{c}{}{$\fth$}
 & \tablehead{1}{c}{}{$\Mfe$}
 & \tablehead{1}{c}{}{$\Mbh$}\\
   \tablehead{1}{c}{}{}
 & \tablehead{1}{c}{}{$[\Msun]$}
 & \tablehead{1}{c}{}{${\rm [km]}$}
 & \tablehead{1}{c}{}{$[10^{51} {\rm ergs~s^{-1}}]$}
 & \tablehead{1}{c}{}{$[10^{51} {\rm ergs}]$}
 & \tablehead{1}{c}{}{${\rm [degrees]}$}
 & \tablehead{1}{c}{}{}
 & \tablehead{1}{c}{}{}
 & \tablehead{1}{c}{}{$[\Msun]$}
 & \tablehead{1}{c}{}{$[\Msun]$}\\
\hline
 A& 1.4 & 900 & 120& 15 & 15 & 100 & 10$^{-3}$ & $2.1\times10^{-1}$ & 9.1 \\
 B& 1.4 & 900 & 1.5& 15 & 15 & 100 & 10$^{-3}$ & $3.9\times10^{-6}$ &16.9 \\
\hline
\end{tabular}
\caption{Jet-induced explosion models.}
\label{tab:2Drelmodel}
\end{table}

\section{RESULTS}
\label{sec:result}

\subsection{Fallback}
\label{sec:fallback}

\begin{figure}
  \includegraphics[height=.2\textheight]{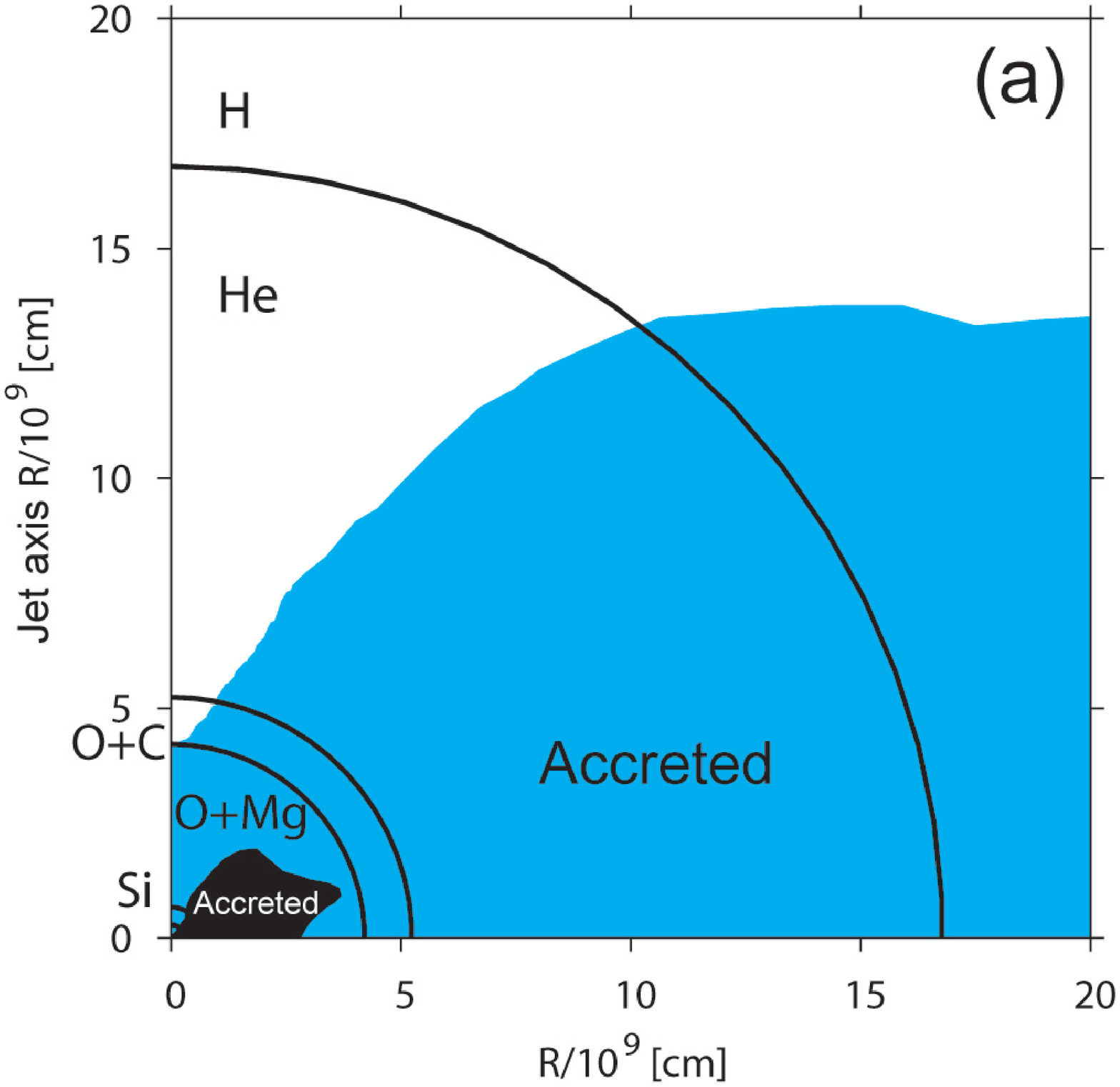}
  \includegraphics[height=.2\textheight]{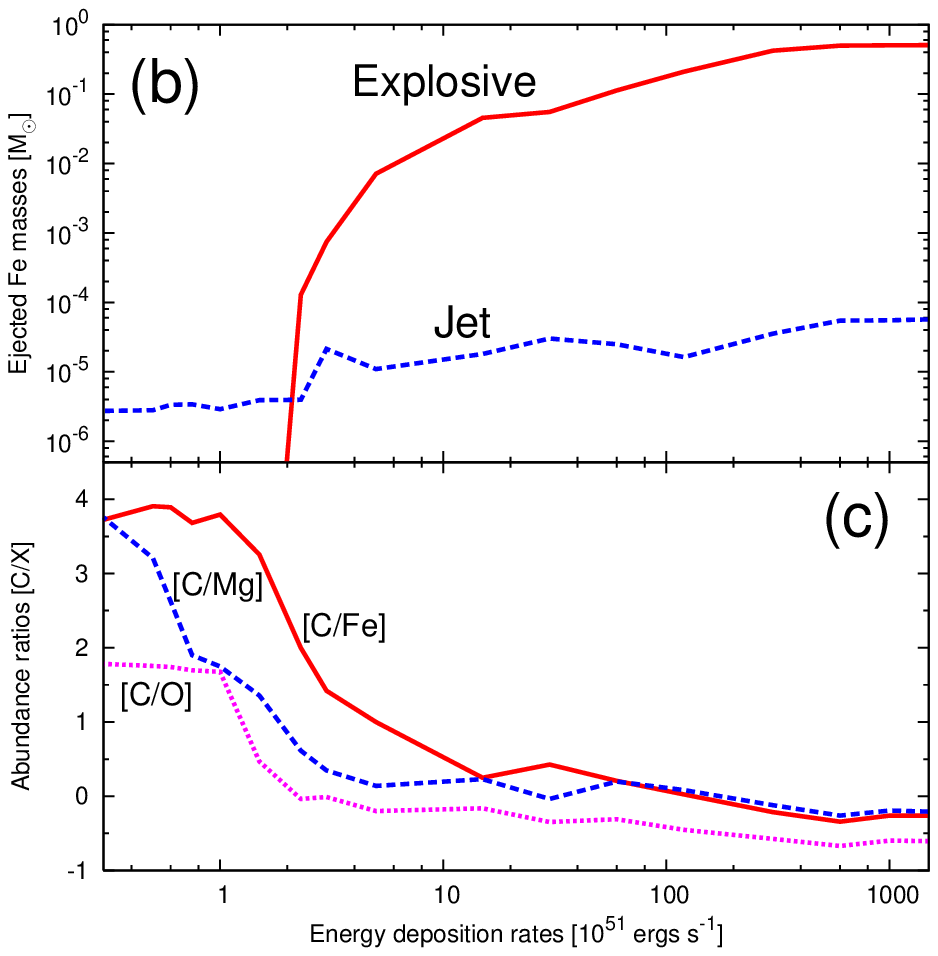}
  \includegraphics[height=.14\textheight]{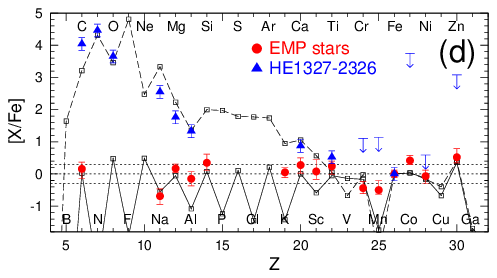}
 \caption{(a) Initial locations of the mass elements which are finally
 accreted for model A ({\it black}) and
 model B ({\it cyan}). The circles represent the boundaries between the layers in the
 progenitor star; the H, He, O+C, O+Mg, and Si layers from the outside. 
(b) Ejected Fe mass ({\it solid line}: 
 explosive nucleosynthesis products, {\it dashed line}: the jet contribution)
 as a function of the energy deposition rate. 
 (c) Dependence of abundance ratios, [C/Fe] ({\it solid line}),
  [C/Mg] ({\it dashed line}), and [C/O] ({\it dotted line}), on the energy
 deposition rate. 
(d) Comparisons of the abundance patterns of metal-poor 
 stars and models. The abundance patterns of the EMP stars
  (\cite{cay04}, {\it circles}) and HE~1327--2326 (\cite{fre05,aok06}, {\it triangles}) are reproduced by
  models with $\Edep=120$ ({\it solid line}) and $\Edep=1.5$ ({\it
  dashed line}).
\label{fig:2Drelfallback}}
\end{figure}

Figures~\ref{fig:2Drelfallback}a shows
``accreted'' regions for models A and B, where the accreted mass
elements initially located in the progenitor \cite{tom07c,tom07d}. The O layer is
separated into the two layers: (1) the O+Mg layer with 
$X({\rm ^{24}Mg})>0.01$ and (2) the O+C layer with $X({\rm ^{12}C})>0.1$. 
The inner matter is ejected along the jet-axis but not along the
equatorial plane. On the other hand, the outer matter is ejected even
along the equatorial plane because the lateral expansion of the shock
terminates the infall.

The accreted mass is larger for lower $\Ed$. This stems from the balance
between the ram pressures of the injecting jet ($P_{\rm jet}$) and the
infalling matter ($P_{\rm fall}$) (\eg \cite{fry03,mae06c}). 
The critical energy deposition rate ($\dot{E}_{\rm dep,cri}$) giving 
$P_{\rm jet}=P_{\rm fall}$ is lower for the outer layer. Thus, the jet
injection with lower $\Ed$ is realized at a later time when the central
remnant becomes more massive. Additionally, the lateral expansion of the
jet is more efficiently suppressed for lower $\Ed$. As a result, the
accreted region is larger and $\Mbh$ is
larger for lower $\Ed$. 

Figures~\ref{fig:2Drelfallback}b and \ref{fig:2Drelfallback}c show the dependence of
$\Mfe$ and the abundance ratios [C/O], [C/Mg], and
[C/Fe] on $\Ed$, respectively \cite{tom07a,tom07c}.
A model with lower $\Ed$ has larger $\Mbh$, higher [C/O], [C/Mg], and
[C/Fe], and smaller $\Mfe$ because of the larger amount of fallback
(Fig.~\ref{fig:2Drelfallback}a).
The larger amount of fallback decreases the ejected mass of the inner core
(\eg Fe, Mg, and O) relative to the ejected mass of the outer layer (\eg C,
Fig.~\ref{fig:2Drelfallback}a). Since O and Mg are synthesized in the
inner layers than C, [C/O] and [C/Mg] are larger for the
larger infall of the O layer. Also, the fallback of the O layer decreases
$\Mfe$ because Fe is mainly synthesized explosively in the Si and O+Mg layers.
Therefore, the variation of $\Ed$ in the jet-induced SN explosions predicts that
the variations of [C/O], [C/Mg], and [C/Fe] corresponds to
the variation of $\Mfe$.

Figure~\ref{fig:2Drelfallback}d shows that the abundance patterns of EMP
stars \cite{cay04} and HE~1327--2326 with [Fe/H]~$\sim-5.6$
\cite{fre05,aok06} are reproduced by models with
$\Edep=120$ and 1.5, respectively (see Table~1 for model
parameters). The model for the EMP stars ejects $\Mfe\sim0.2\Msun$. On
the other hand, the model for HE~1327--2326 ejects 
$\Mfe\sim4\times 10^{-6}\Msun$.

\subsection{Abundance distribution}
\label{sec:angledep}

\begin{figure}
  \includegraphics[height=.32\textheight]{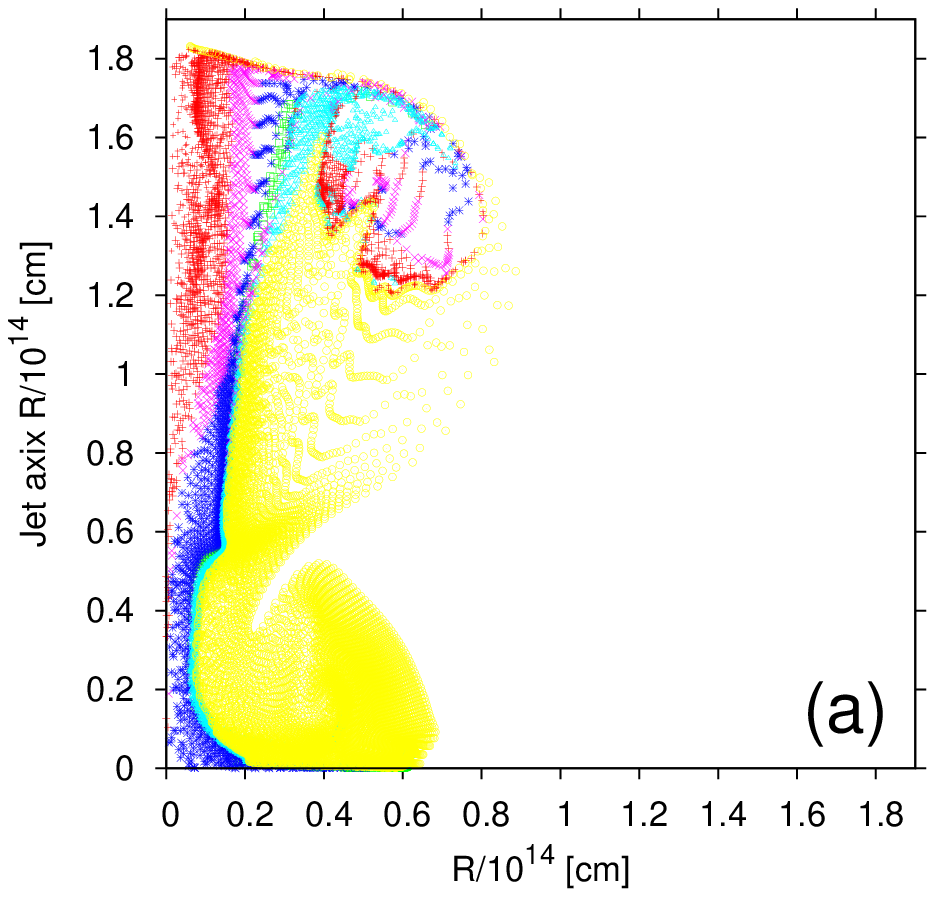}
  \includegraphics[height=.32\textheight]{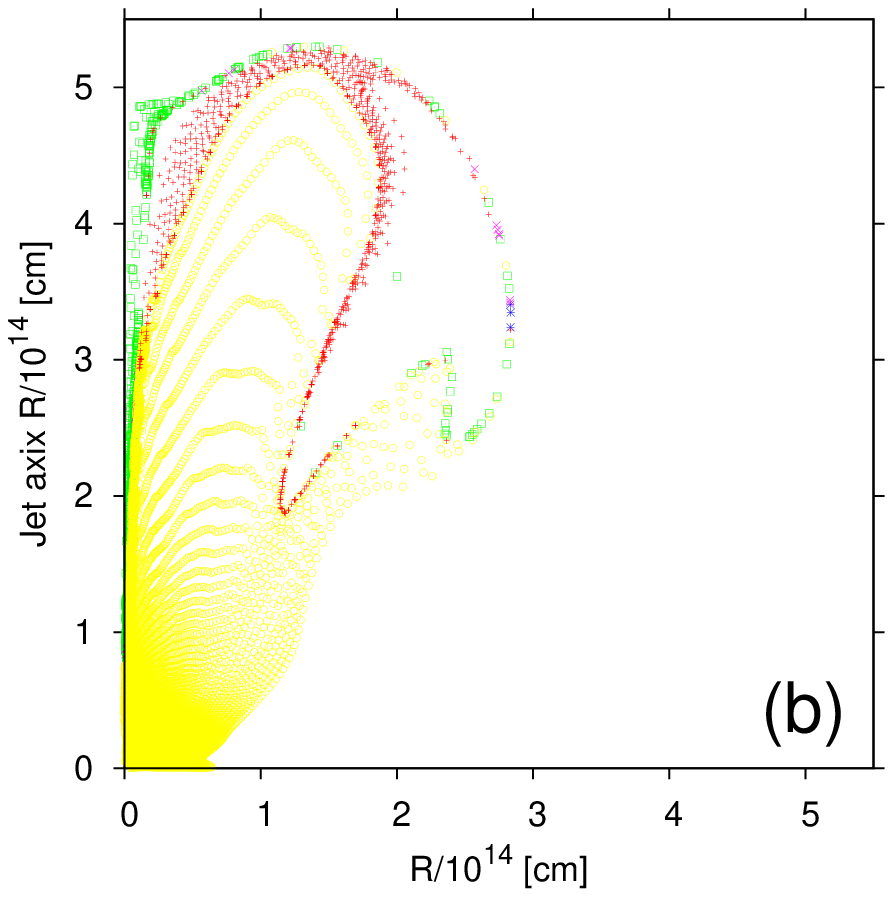}
 \caption{Positions of the mass elements at $t=10^5$~s
 for (a) model A and (b) model B. Symbols of the marks
 represents the abundance of the mass element (H: {\it yellow circles}, He: {\it
 cyan triangles}, O+C: {\it green squares}, O+Mg: {\it blue asterisks},
  Si: {\it magenta crosses}, and Fe: {\it red
 pluses}). Size of the marks represents the origin of the mass element
 (the jet: {\it small}, and the shocked stellar mantle: {\it
  large}).\label{fig:2Drelej}}
\end{figure}

Figures~\ref{fig:2Drelej}a and \ref{fig:2Drelej}b show the abundance
distributions at $t=10^5$~s for models A and
B \cite{tom07d}. We classify the mass elements by their abundances as follows:
(1) Fe with $X({\rm ^{56}Ni})>0.04$, (2) Si with $X({\rm ^{28}Si})>0.08$, 
(3) O+Mg with $X({\rm ^{16}O})>0.6$ and $X({\rm ^{24}Mg})>0.01$, 
(4) O+C with $X({\rm ^{16}O})>0.6$ and $X({\rm ^{12}C})>0.1$, 
(5) He with $X({\rm ^{4}He})>0.7$, and (6) H with $X({\rm ^{1}H})>0.3$.
If a mass element satisfies two or more conditions, the mass element
is classified into the class with the smallest number.

\begin{figure}
  \includegraphics[height=.2\textheight]{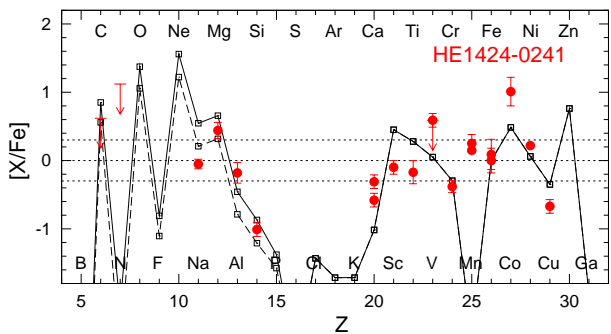}
 \caption{Comparison between the
 abundance pattern of HE~1424--0241 ({\it filled circles}) and the
 angle-delimited yields of model A for $30^\circ\leq\theta<40^\circ$ ({\it
 solid line}) and $30^\circ\leq\theta<35^\circ$ ({\it dashed line}). \label{fig:2DrelAB}}
\end{figure}

The abundance distribution and thus the composition of the ejecta depend
on the direction. In model A, the O+Mg, O+C, He, and H mass elements
locate in the all direction. On the other hand, most of the Fe and Si
mass elements locate at $\theta<10^\circ$ and stratify in this order
from the jet axis and a part of them locate at
$15^\circ<\theta<35^\circ$. Interestingly, the Fe mass elements surround
the Si mass elements at $15^\circ<\theta<35^\circ$. In model B, most of
the O+C and He mass elements locate at $\theta<3^\circ$, while the Fe
mass elements expand laterally up to
$\theta\sim50^\circ$ and the H mass elements are distributed in the all
directions. The lateral expansion of the Fe mass elements in
models A and B are led by the collision with the stellar mantle and
the internal pressure of the jet.

A very peculiar, Si-deficient, metal-poor star HE~1424--0241
was observed \cite{coh06}. Its abundance pattern with high [Mg/Si] ($\sim1.4$)
and normal [Mg/Fe] ($\sim0.4$) is difficult to be reproduced by
previous SN models. This is because 
$\log\{[X(^{24}{\rm Mg})/X(^{28}{\rm Si})]/[X(^{24}{\rm Mg})/X(^{28}{\rm Si})]_\odot\}\lsim1.6$ 
is realized in the O+Mg layer at the presupernova stage (\eg
\cite{woo95,ume05}). Thus, in order to reproduce
the abundance pattern of HE~1424--0241, it is required to consist of
explosively-synthesized Fe but not explosively-synthesized Si.

The angle-delimited yield may possibly explain high
[Mg/Si] and normal [Mg/Fe]. Figure~\ref{fig:2DrelAB}
shows that the yields integrated over $30^\circ\leq\theta<40^\circ$ and
$30^\circ\leq\theta<35^\circ$ of model A reproduce the abundance pattern of HE~1424--0241. 
The yield consist of Mg in the inner region and Fe in the outer region
(Fig.~\ref{fig:2Drelej}a).
Although there are some elements to be improved, the elusive feature of
HE~1424--0241, high [Mg/Si] and normal [Mg/Fe], could be explained. The elusive feature can be
realized with a model satisfying the following conditions: (a) the Fe mass elements
penetrate the stellar mantle (\ie the duration of the jet injection is
long) and (b) the O+Mg mass elements are ejected in all directions (\ie
$\Ed$ is high).

\section{Conclusions and Discussion}
\label{sec:summary}

We perform two-dimensional hydrodynamical and nucleosynthesis
calculations of the jet-induced explosions of a Pop III $40\Msun$ star
and show two jet-induced explosion models A and B as summarized in
Table~\ref{tab:2Drelmodel}.
We have shown that (1) the yields of the
explosions with high $\Ed$ explain the
abundances of the EMP stars, and (2) the explosions with low $\Ed$
are responsible for the
formation of HE~1327--2326.

(1) {\bf Fallback}: The dynamics and the abundance distributions depend sensitively on
$\Ed$. The explosion with lower $\Ed$
leads to a larger amount of fallback, and consequently smaller $\Mfe$
and higher [C/O],
[C/Mg], and [C/Fe]. Such dependences of [C/Fe] and $\Mfe$ on $\Ed$
predict that the higher [C/Fe] tends to be realized for
lower [Fe/H]. Note, however, the formation of star with low [C/Fe] and
[Fe/H] is possible because [Fe/H] depends not only on $\Mfe$ but also on
the swept-up H mass, \ie the interaction between the SN ejecta and
interstellar matter (ISM) (\eg \cite{cio88}).

(2) {\bf Angular dependence}: We present the aspherical abundance distributions and investigate the
angular dependence of the yield. The angle-delimited
yield could reproduce the extremely peculiar abundance pattern of HE~1424--0241.
However, we note that the angle-delimited yield depends strongly on which mass elements are
included into the integration. This would be determined
by the abundance mixing in the SN ejecta, the interaction between the SN
ejecta and ISM, and the region where the
next-generation star takes in the metal-enriched gas. It is
necessary to calculate three-dimensional evolution of the supernova
remnant (\eg \cite{nak00}).

\vspace{.5cm}

\begin{theacknowledgments}
\noindent This work has been supported in part by World Premier International
Research Center Initiative (WPI Initiative), MEXT, Japan, and by the
Grant-in-Aid for Scientific Research of the JSPS (18104003, 18540231,
20244035, 20540226) and MEXT (19047004, 20040004).
N.T. is supported through the JSPS (Japan Society
for the Promotion of Science) Research Fellowship for Young Scientists.
\end{theacknowledgments}



\bibliographystyle{aipproc}   

\bibliography{sample}

\IfFileExists{\jobname.bbl}{}
 {\typeout{}
  \typeout{******************************************}
  \typeout{** Please run "bibtex \jobname" to obtain}
  \typeout{** the bibliography and then re-run LaTeX}
  \typeout{** twice to fix the references!}
  \typeout{******************************************}
  \typeout{}
 }


\end{document}